\documentstyle[12pt]{article}

\normalbaselines

\newcommand{\be}{\begin{equation}}
\newcommand{\ee}{\end{equation}}
\newcommand{\ba}{\begin{eqnarray}}
\newcommand{\ea}{\end{eqnarray}}
\newcommand{\baa}{\begin{eqnarray*}}
\newcommand{\eaa}{\end{eqnarray*}}
\newcommand{\bb}{\begin{thebibliography}}
\newcommand{\eb}{\end{thebibliography}}

\newcommand{\bi}[1]{\bibitem{#1}}
\newcommand{\lab}[1]{\label{#1}}
\newcommand{\re}[1]{(\ref{#1})}

\newcounter{my}
\newcommand{\he}%
   {\stepcounter{equation}\setcounter{my}%
   {\value{equation}}\setcounter{equation}1%
   }%
\newcommand{\she}%
   {\setcounter{equation}{\value{my}}%
    }%

\begin{document}

\vspace*{2mm}

\begin{center}

{\Large \bf  Does There Really Exist the Problem\\3mm of the Dark Matter
in Spiral Galaxies?}

\vspace{7mm}

\bigskip\medskip

{\large \bf A.E.Filippov\footnote{E-mail: medvedev@host.dipt.donetsk.ua
}} and {\large \bf
A.S.Zhedanov\footnote{E-mail: zhedanov@host.dipt.donetsk.ua}}\\

\vspace{2mm}
{\it Donetsk Physical Technical
                   Institute National Academy of Sciences 340114, Donetsk,
                     Ukraine}\\

\end{center}

\vspace{2mm}

\begin{abstract}
A simple model for the dust media describing evolution of
the system like spiral galaxy is considered. In contrast to previous
considerations we show that the initial density fold should be quasi-one-
dimensional (bar-like) instead of disc-like. The disc component of the
galaxies appears only during the evolution. The model naturally reproduces
some essential features of the galaxies. In particular, it reproduces all
the observed typical forms of the rotation curves for the spiral galaxies
with a characteristic minimum and plateau. It appears that the plateau
corresponds to escaping the matter (external spiral arms, due to initial
conditions, have too large velocities to be confined by the gravitational
field of the galaxy).  Such a scenario of the galaxy evolution leads to the
conclusion that the hypothesis of the dark matter is not necessary (at least,
for the spiral galaxies).
\vspace{2cm}


\vspace{2cm}

PACS index: 95.35.+d, 98.10.+z, 98.62.Ai, 98.62.Dm

\end{abstract}

\newpage

\newpage

\section{Introduction}

A so-called "virial paradox" arises naturally from observations of stellar
systems, galaxis etc. If we suppose that a given system moves  stationary
then the virial theorem should be valid \cite{vv}
\be
2T = - U,
\lab{vt} \ee
where $T$ is an averaged kinetic energy  of  the  system  (excluding center of
mass motion) and U is an averaged value of its potential (gravitational)
energy.  Generally speaking, these averages  are assumed over a time.
But, in the astronomy  we  can  fix  the ansemble averaged values only. So,
usually peculiar  "astronomic" ergodic hypothesis is accepted.  In  this case
averaged  kinetic energy may be written in the form
\be
2T = M \langle v^2 \rangle ,
\lab{kin} \ee
where $M$ is a total mass of the system and $\langle v^2 \rangle$ is
averaged squared velocity.  The  values  $U = - GM^2/R$  and  $\langle
v^2 \rangle$ may be estimated from  the observations (here $R$ is mean radius
of the ansemble).  It should be noted that in reality we find  a component of
the velocity directed to the observer, because  the only Doppler effect  may
be used to calculate this   value independently.  Using \re{vt} and \re{kin}
we find an estimation for total mass of the system:
\be
M_V = R \langle v^2 \rangle /2G
\lab{virmass} \ee
The value $M_V$ obtained from \re{virmass} is called "virial mass" of the
system. Of course, there is uncertainty in determining $M_V$
connected with estimation  of the mean radius  of the system $R$.
Nevertheless numerous observation yield the value of $M_V$ which appears to
be much more gretaer than the mass $M_o$ estimated from stars, dust and
other types of visible matter. Ordinary, $M_V/M_o \sim 10$ but there are
systems where this ratio may be of order of $10^2$. This  fact  is  the
"virial paradox" mentioned above.  It is interesting  to  note  a  specific
"scale invariance" property. The paradox takes place starting from the star
system forming a nearest vicinity of the Sun  \cite{mar}  up  to all highest
scales (namely for the star clusters  in our Galaxy, the galaxies by
themselves and for the galaxy clusters).  The same conclusion may be found
studying a kinematics of  the spiral galaxies.  Indeed, let  us  suppose
that  the main  masses forming the galaxy rotate stationary around the galaxy
center.  So, their orbits are close to the circles. If an edge of the galaxy
is visible we can plot a so-called rotation curve. It is a dependence of the
star velocity component directed to  the observer  on  the distance between
this star and galaxy center. Best known rotation curve for the M31  galaxy
(Andromeda  nebula) is  strongly  non- monotonous and has a "strange" plateau
on galaxy periphery (see $de$ fragment on the figure 22 in book \cite{vv}, for
example). Central parts of the curve (like $ab$ and $bc$ on the same figure)
may be  invisible for some other galaxies due to large  distance  or  dust
presence near their center.  But, the plateaus are  characteristic  features
for the other spiral galaxies too.  This fact forces astronomers to assert
on existence of some additional  in- visible masses ("dark matter") in
the galaxies. Indeed,  if  we suppose that the main mass of galaxy is
localized near its center an obvious law
\be
v^2 \propto   1/r
\lab{vel} \ee
takes place. This law
contradicts to the $de$ part of  the  rotation curve. We may treat $de$ line
as a horizontal one approximately.  In this case it is necessary to have an
additional  mass  spherically distributed with a density $\rho \propto r$  .
However, observation  do  not give other evidences of its existence in spiral
galaxies.  Hypothesis of the dark matter is accepted by majority of the
astrophysicists (see, however, \cite{gott}, where it was shown that at least
for some spiral galaxies the hypothesis of massive galo of dark matter
contradicts to observed motion of the galaxy satellites). Theoretically,
however, two other variants of a solution of this problem have been proposed.
They are:  1) possible violation of the gravitation law at large (galactical)
scales; relativistic variant of this hypothesis was analyzed recently in the
work \cite{prl}, where  authors conclude that this supposition does not
effective; 2) V.A.Ambartsumyan hypothesis  \cite{va} that  the  star
clusters, galaxies and their clusters are strongly  unstable (flying  away)
objects which arose as the results of an explosion  of  some protostar
substance.  This hypothesis takes  away the virial paradox, obviously.  But,
it meets with many essential difficulties connected  with the  physical
interpretation of the  protostar objects, the reasons  of their explosion
with extremely high energy, etc.  (see  critical comments on  this conception
in Ref.  \cite{zn}).  From our point of view, a general idea of the
instability  of star (and more generally - self-gravitating) systems  is a
most fruitful one in V.A.Ambartsumyan hypothesis.  In  turn, the instability
should not be related to any exotic hypothesis of  the "explosion", etc.
Conclusion about run  away  of  the peripheral parts of the galaxies does not
require any supposition  outside standard physical theory. Moreover, we
intend to show that such  a behaviour is a {\it necessary  scenario}  at
general  form of initial conditions which are characteristic ones at galaxy
formation  from the dust medium.

Recently, new kinetic approach to study  of  the  dust  medium
evolution has been proposed by one of authors \cite{f1}, \cite{f2}. Below it
will be shown that using this approach virial  paradox  may  be  solved quite
naturally. In short our solution will be based on  the  fact (stably
observable in numerical simulations using the model \cite{f1}, \cite{f2} that
{\it a part} of the substance in the galaxy really  runs  away.  It means
that the virial theorem can not be utilized here in the form of \re{vt}.
Moreover, our numerical experiments provide  extremely  strong agreement
between calculated and observed "rotation curves" for spiral galaxies.
Really, as it will be clear below,these curves do not describe rotation but
instead correspond to some complicated kinetic processes in the system.  They
are running away with "twisting out" for peripheral branches and compressing
with "twisting in" for internal regions of system. As a result, peripheral
plateau and other features  of the rotation curves arise naturally and does
not require latent mass hypothesis to be explained.

\section{Model and results}

A simple  kinetic  model  of  dust  media  evolution  has  been
proposed in \cite{f1},\cite{f2}. Let us repeat briefly the main ideas  used
and some general features of the model which  may  be  interesting  in
context of present article.

Numerical study of three (and more) body problem  shows  that
the dynamic chaos is a typical behaviour of the system \cite{zas}, \cite{f3}.
This behaviour even leads to  instability  of  the  system  with  small
number of bodies. Chaotic  behaviour  allows  use  a   statistical
approach to many body problem. Dynamic equations with random noise
source may be used instead of  an  exact  ones.  For  sufficiently
large system simplest form of $\delta$-correlated noise may be  used.  In
its turn, some of the interactions with regularly  moving  objects
should be accounted in the equation directly \cite{f1}, \cite{f2}.

Formal use of the noise leads to  energy  production  in  the
system. The equation of motion should have a dissipative  term  to
restore the energy conservation.  Dust  evolution  leads  to  some
energy dissipation. Dissipative effect in the equation  should  be
taken larger than effect  of  the  noise  to  simulate  this  real
dissipation. Below, we return to this question.

Above notes lead to the equation
\be
d^2 {\bf r}_k /dt^2  = - \gamma d{\bf r}_k /dt - {\bf f}(t,r_k ) - G \sum_j
{m_j \frac{{\bf r}_k-{\bf R}_j} {|{\bf r}_k-{\bf R}_j|^3}}
\lab{eq} \ee

where vectors ${\bf r}_k = (x_k ,y_k ,z_k )$ are the positions of  the
particles, ${\bf R}_j$  are the positions of the  regularly  moving  objects
with the masses $m_j$  respectively and (or) the positions of the mass
centers of the dust clouds. White $\delta$-correlated noise
\be
\langle {\bf f}(t,{\bf r}_k ) \rangle= 0;
\langle {\bf f}(t^{\prime},{\bf r}^{\prime}_k)f(t,{\bf r}_k ) \rangle=
D \delta(t-t^{\prime}) \delta({\bf r}-{\bf r}^{\prime}))
\lab{corr} \ee
is taken for a simplicity.  In this note we concentrate  on  typical
scenarios of the galaxy
evolution.  Simplicity  of  model  allows  simulate both:  formation of the
galaxies from trial  density  folds  and  their collisions.  In this case the
${\bf R}_j$ values are the mass centers of the galaxies.  Limiting ourselves
by one galaxy arising and calculating at each step its mass center:
\be
{\bf R} =\frac{\sum_k{ m_k {\bf r}_k}}{\sum_k{m_k}}
\left( = \sum{ {\bf r}_k} / N  \quad \mbox{at} \quad m_k =1  \right)
\lab{cm} \ee

one has a very compact equation
\be
d^2 {\bf r}_k /dt^2 = - \gamma d{\bf r}_k /dt - {\bf f}(t,{\bf r}_k ) -
G\frac{{\bf r}_k -{\bf R}}{|{\bf r}_k -{\bf R}|^3}.
\lab{comp} \ee
These equations should be completed by initial  conditions.  Fold
catastrophe formation in a dust medium  density is  known  as  a probable
initial stage of galaxy formation \cite{trub},\cite{cher},\cite{dor} and
namely such dust configurations were used as initial conditions in
\cite{{f1},{f2}}.  They give time transformations of the initial fold-like
configuration which are very close to the typically observed.  Few
intermediate states obtained at numerical simulations are shown on the Fig.1.
Dust beams in initial fold do not parallel in  general case.  It  leads  to a
rotation  of  the system.A dispersion  of the velocities leads to a formation
of the typical structure depicted.  As a rule it tends to increase galaxy
volume. It is seen directly from the Fig.1 that final galaxy-like  structure
is bigger  then starting fold. In turn, dissipation effect tends  to decrease
phase volume of the system.  Dissipation  has  maximum  at  maximal
velocities near mass center. Last fact even  makes  a  model  more
self-consistent, because system has maximal density in this region which
should favor to the dissipation too. As result the particles can not  return
from  vicinity  of mass  center  to  more large distances \cite{ll} and total
system forms galaxy-like picture.  According to general thermodynamic
principles (see \cite{ksf}  and references there) temp of relaxation tends to
its minimum.  System goes to a stationary dissipative attractor. In  works
\cite{f1,f2} this fact was treated as the main reason of stable formation of
galaxy- like structures from quite arbitrary configuration of dust in  the
space. Lowest relaxation corresponds to  a motion  along  special saddle-like
trajectory (so-called "large river"). This trajectory separates two different
kinds of flow lines.  First  of  them  are lines having more high  velocities
then $v_c$ ,  going outside  the separatrix and second ones with $v < v_c$
shooting to the center (here $v_c$ denotes the velocity on circular
orbit where the tempo of relaxation is minimal). Let us plot a
specific phase pattern of the system on the coordinates $K = v^2 /2;
|U|=1/r$. Fig.2 presents a developed intermediate stage  of  the evolution
shown on above coordinates.  For convenience  two characteristic straight
lines corresponding to the energy balance $K = |U|$ and to the virial
relation $2K = |U|$ (lines  $A$ and  $B$ respectively)  are shown also. Three
different families of the "star population"  may be separated here.

1) Above the line $B$ the  points  running  away  are  located.
Their configuration preserves partially  a  structure  of  initial
configuration. On the figure presented it has been  taken  in  the
form of  long  density  fold  with  tangential  velocities  slowly
decreasing with distance from the center of  structure.  Evolution
shifts this part of density distribution to more small $|U|$  and  $K$
values. It corresponds obviously to a run  away  of  the  external
branches. They are quite visible on the peripheral regions  of
the insert to the Fig.2. reproducing the galaxy structure in  real
space corresponding to given phase portrait.

2) In  right  hand  side  of  the  Fig.2  the  points  moving
relatively close to the center are shown. They correspond  to  the
developed spiral-like structure shown on the insert. It should  be
noted  specially  that  this  structure  is  not  a  real   spiral
trajectory of the points. It is  a  current  distribution  of  the
particles only. In reality the points  forming  this  spiral  move
around the center along (quasi-) elliptic mechanical trajectories.
Numerical simulation gives "paradoxical" evolution of this central
spiral. It seems like generating and untwisting from  the  center,
but averaged radius of this central part of massive decreases with
time.

Internal branches are more  strongly  twisted  than  external
ones. Being the parts of  the  distorted  mechanical  trajectories
they sometimes intersect forming pictures which seems like  spiral
galaxies with bars (or bridges).

The phase portrait (on  $K,|U|$  coordinates)  of  the  central
spiral structure tends at $t \to \infty$  to a straight line parallel to the
line $A$. It corresponds to a simple conservation law $K + U = const$
due to balance between noise and relaxation.

3) One of the most interesting  region  on  the  Fig.2  is  a
fragment  of  the  distribution  located  near  the  line  $B$.   It
corresponds to the vicinity of the  separatix  where  points  move
with velocities close to the  $v_c$ .  At  exactly  circular  velocity
virial theorem is automatically satisfied for each point and $2K_j  =
|U_j|$. Projection to the plane ($K,|U|$) gives for them a line  close
and parallel to the  line  $B$.  Initial  stages  of  the  evolution
transform trial fold configuration to the vicinity of this line as
to the dissipative attractor. At this initial stage system relaxes
along the line $B$. This relaxation produces first, second and other
internal branches of spiral starting from the line B and  forms
a developed picture shown on Fig.2.

At relatively late stages  of  the  process  all  points  are
strongly separated into 1)-st and 2)-nd classes. Vicinity  of  the
separatrix (class 3)) becomes very impoverished. Galaxy takes  the
form shown on the Fig.3 (in two projections).  Visually  it  seems
like elliptic or spiral (with strongly  twisted  branches)  galaxy
having couple of satellites jointed with the central  "galaxy"  by
two very thin channels. These configurations  were  found  by  the
astronomers quite often producing some artificial explanations  of
the reasons for the satellites formation. In its turn, numerically
calculated final picture like Fig.3 is produced by the system as a
very  typical  too.  It  seems  like  very   natural   theoretical
explanation of  such  galactic  configurations  in  the  frame  of
present kinetic model.

Rotation curve calculated should reflect  the  separation  of
the "star population" into three  families  also.  Doppler  effect
allows to plot such a curve for real galaxies using one  direction
(relatively accidental) of the velocity only. Model gives formally
more complete information  to  plot  a   map  of  the
velocities and compare it with the maps partially  available  from
astrophysical literature \cite{vv}. For example, we may compare
analogous curve calculated with a
plot of the $z$-component of the velocity (orthogonal to the  galaxy
plane). It is shown on Fig.4. More  regular  (and  representative)
characteristics is a mean square of the velocity  which  shown  on
Fig.5.

It is seen directly  that  $z$-component  of  the  velocity  is
generated by its small perturbations near the galactic center.  It
produces z-component  of  the  coordinate  with  maximum  slightly
shifted from zero r. At given time moment its value depends  on  a
structure of initial dust configuration and on intensity of  noise
(mutual interactions). Combining them one may produce very natural
galaxy-like  pictures.  For  example,  already   mentioned   Fig.1
displaces  a  picture  where  two  different  components  of   the
population are  accounted.  Bold  points  here  simulate  strongly
interacting  "stars"  from  relatively  compact  central  part  of
starting fold and small points correspond to a dust  component  of
the system respectively.

The Figs.4 and 5 are plotted for different  initial  velocity
distributions (see the sequences  of  the  inserts  there  showing
different time steps of the evolution). It  allows  to  see  both:
universal features of the process in central part  of  the  system
and some memory about initial conditions conserved on the external
branches. All these different structures of  rotation  curves  are
observed for different real galaxies. For example,  insert  c)  on
Fig.4 demonstrates  unexpectedly  good  coincidence  with  already
mentioned data for Andromeda galaxy (see \cite{vv}). The insert a) on Fig.5
corresponds to the typical rotation curve for spiral galaxies, whereas
the inserts b),c) and d) on Fig.5 demonstrate very close correspondence
with observed rotation curves for the galaxies:  NGC 7541, NGC 2998 and NGC
801 respectively \cite{bos}, \cite{rubin}. Thin line on the main plot of
Fig.5 shows an integral mass distribution as function of $r$, which is also
close to the estimated from the observations.

Analyzing  numerical  data  one  may  extract  some  averaged
information about the system. Most interesting for us  here  is  a
relation between $2K$ and $|U|$ used in virial theorem. Fig.6 shows  a
typical time dependence for this relation in the  frame  of  model
(curve $A$). It  may  be  proved  directly  that  for  most  of  the
scenarios leading to the typical galaxy-like  space  distributions
initial ratio $2K/|U|$ should be much more larger than $1$
required by the
virial theorem for stationary moving objects.  Our numerical experiments give
the typical relations: $2K/|U|  =  2\div 10$.  Value of $2K/|U|$ decreases
with the time and at  $t \to \infty$  tends  to limit value which is  defined
by a balance between  noise  and relaxation in the model. For
stationary object  this relation is expected to be $2K/|U| =1$ exactly. It
gives good self- consistent criterion used by us to fix a relation between g
and  D constants in the Eqs. (2.1) and (2.2).  Moreover,  it  puts  a
natural  limit  of  the  model  validity,   because   a   physical
dissipation can not be accounted after reaching of this limit.

To complete the discussion of the Fig.6 let us note that  the
evolution of the perturbations of the $z$-component of the  velocity
already discussed above leads to growth of  averaged  $z$-coordinate
of the particles in central part of the system.  Curve $B$  on  the
Fig.6 shows time dependence of the ratio between averaged $(x,y)$ and
$z$ coordinates in the central subsystem. It is  interesting  to
note that the curve has a typical maximum at relatively  small  $t$.
It is supported by direct observations of the system evolution. Indeed, at
small $t$ large  angular momentum of initial dust configuration leads to
increasing the "planarity" of the system. This process affects  internal
parts of  the  system  too.  But then, after some evolution,  growth   of
z-coordinate leads to an isotropization of  the  central  part  of
distribution producing an elliptic (or toroidal) galaxy as result.

 \section{Summary}
It is found that different observed features of the  galaxies
may be explained without artificial suppositions  like:  formation
of trial dust disk in the space; its rotation  like  solid  system
and other additional hypothesis. The  only  density  folds  stably
generated by the  dust randomly moving in the space are sufficient
sources for galaxy-like structure arise.

We used a simple model of dust media  evolution  which  is  a
combination of mechanical and statphysical approaches. It accounts
a nonstationarity of the systems and describes  an  evolution  the
galaxy-like  systems  in  the  terms  of   their   relaxation   to
dissipative attractors. This model reproduces some features of the
galaxies quite naturally. These  features  may  be  summarized  as
follows.

1. Arbitrary density fold evolves to the structures close  to
the observed for real galaxies \cite{vv}, \cite{arp}, \cite{vac}.

2. Phase  portrait  of  the  system  is  transformed  to  the
attractor form providing with dissipation minimum.

3. At least  three  characteristic  families  of  the  moving
points in the galaxy structure may be separated:

\begin{quote}(i). Central part of the  system  having  densely  parked
and strongly twisted  branches.  This  subsystem  tends  to  the elliptic
      form at large stages of the evolution;

\end{quote}

\begin{quote}(ii). Peripheral part  of  the  system  with  (typically)  two
      branches running away. These branches are twisted much  more
      weakly than central ones;

\end{quote}

\begin{quote}(ii). Intermediate  region  corresponding  to  a
      separatrix regime. Particles move here with a lowest velocity providing
      with a minimum of dissipation. With the time all points tend
      to be strongly separated into internal and external classes.
      Vicinity of the separatrix  becomes  very  impoverished  and
      system takes (typical) form of the central galaxy  with  two
      satellites.
\end{quote}

4. Model reproduces typical forms  of  rotation  curves known
from the observations. In particular, standard separation  of  the
system into three regions reproduces  a  rotation  curve  with  a
characteristic minimum in the separatrix region. It should be stressed that
the term "rotation curve" does not correspond to the real motion of galaxy:
only particles near the separatrix can be considered as rotating around the
center of mass. This separatrix corresponds to the minimum on the observed
"rotation curve".

5. For the scenarios leading to  the  galaxy-like  structures
the initial relation between doubled kinetic and potential  energy
$2K/|U|$ should be larger than the condition $2K/|U| = 1$ required  by
the virial theorem for stationary moving  objects.  Our  numerical
experiments give the typical relations: $2K/|U| = 2\div 10$. This result
quite  naturally  takes  away  a  virial   paradox   without   any
supposition about dark matter or other additional hypothesis.

6. The spiral arms structure arises quite naturally in our model as dynamical
attractors of the particles moving in Newtonian gravitational potential.
The "external" and "internal" arms have different origin and different
evolution:  the first one are escaping from galaxy, whereas the second one
become more twisted and dense during the time. Note that we need not any
(rather artifical) hypothesis, like density waves, for explaining the oprigin
of spiral arms.

7.It is clear that generic initial  density
fold should be {\it quasi-one-dimensional} (bar-like), in contrast to
the commonly accepted opinion that initial fold is disc-like.  This leads to
the conclusion that disc component is not necessary for spiral galaxies.
Indeed there are pecular non-complanar galaxies with arms but without disc
(see bright examples of such "exotic"
non-complanar spiral galaxies in \cite{vv}). In our approach the
non-complanarity is explained merely by appropriate choosing of the
initial velocities in the fold (clearly, these velocities need not be
complanar, in general).  In turn, absence of initial disc allows to resolves
the problem of the origin of galaxy rotation (in standard scenarios with
initial disc this problem is crucial).

8. A wide spectrum of observed forms of spiral galaxies is explained by:
\begin{quote} (i) difference of initial conditions for the density fold;

(ii) time evolution: the form changes during the evolution. Moreover the
spiral galaxies are {\it essentially non-stationar} objects
contrary to commonly accepted point of view.
\end{quote}

9. We see from the scenario that usual alternative for
resolving the viral paradox - either completely bound, stationar system, or
completely unbound, explosing one - is not correct. The situation may be much
more complicated: coexistence of two components - bound and unbound. What is
more important is essential non-stationarity of the matter distribution in
the galaxy leading to the conclusion that in general, the virial theorem
should not be valid for gravitational systems. It is intersting to note that
recently the same conclusion was made studying the model of three interacted
galaxies \cite{cher1}.

\section*{Acknowledgments}
One of the authors (A.Zh.) is very indebted to Institute of Theoretical
and Experimental Physics (Moscow) for hospitality. The work is
partially supported by ISF grants K58100 (A.F.) and U9E000 (A.Zh.).

\newpage
\begin{center}
                        Captions to figures
\end{center}

Fig.1 (a-d) Typical stages of the galaxy evolution at $\gamma = 10^{-2}  ,  G
= 0.15, D = 10^{-4}$  . Two different types of the points are  described
in the main text. On the insert a projection to the  galaxy  plane
orthogonal is presented for the last of the shown stages.

Fig.2 Phase pattern of the galaxy-like distribution shown  on  the
$K,|U|$ space. Lines A and B correspond to the equalities $K =  |U|$
and $2K = |U|$  respectively.  The  same  galaxy-like  structure  in
physical $x,y,(z=0)$ space is shown on the insert to the picture.

Fig.3 Two projections (orthogonal to the $z=0$ plane and  close
this plane respectively) of the configuration  having
naturally generated satellites.

Fig.4 Transformations of rotation curve (upper points) with a time
shown as a projection of the numerical data on the $v_x$  and $r$  coordinates.
Down points correspond to the same evolution of the $v_z (r)$
dependence. Series of the inserts show different  time  stages  of
the evolution of one initial configuration. Bold line on the  main
picture presents an average v  at each given value of radius r.

Fig.5 The same as it  is  on  the  Fig.4,  but  for  the  absolute
velocity $v = \sqrt{v_x^2  +v_y^2  +v_z^2}$    instead of the $v_x$ .

Fig.6 Time dependencies of the relation $2K = |U|$ (line  A)  and
the  ratio  between   averaged  $(x,y)$   and   $z$   coordinates
(ellipticity) in the central region of the system (line B).

Fig.7 Different "peculiar" forms of the galaxies obtained  in  the
frame of model. Near each calculated  picture  the  real  galaxies
having close qualitative forms are given.

\newpage

\bb{99}

\bi{vv}  B.A.Vorontsov-Veliaminov,  Extra-Galaxy   astronomy,  2-nd
edition. Moscow, "Nauka", 1978. (in Russian).

\bi{mar} D.Ya.Martynov, General astrophysics.,  4-th ed.,  Moscow,
"Nauka", 1988. (in Russian).

\bi{gott} S.T.Gottesman and J.H.Hunter, Astrophys.J., {\bf 200} (1982), 65.

\bi{prl} V.V.Zhytnikov and  J.M.Nester,  Phys.Rev.Lett.,  {\bf 73}  (1994),
2950.

\bi{va} Problems of  modern  kosmogony,  Moscow,  "Nauka",  1978.  (in
Russian).

\bi{zn} Ya.B.Zeldovich  and I.D.Novikov. Structure  and  evolution  of
the Universe, Moscow, "Nauka", 1978. (in Russian).

\bi{f1} A.E.Filippov, Phys.Lett. {\bf A189} (1994), 361.

\bi{f2} A.E.Filippov, Teor.Mat.Phys. {\bf 103} (1995), 161 (in Russian).

\bi{zas} Zaslavskii G.M., Sagdeev R.Z.  Introduction  to  the  nonlinear
science. Moscow, "Nauka", 1988 (in Russian).

\bi{f3} Filippov A.E. (Sov.Phys.) Teor.Mat.Phys., {\bf 94}, (1993) 325 (in
Russian).

\bi{trub} Trubnikov B.A. Pis'ma v ZhETF, {\bf 47} (1988) 365 (in Russian).

\bi{cher} Chernin A.D.  Pis'ma v ZhETF, {\bf 11} (1970) 317 (in Russian).

\bi{dor} Doroshkevich A.G. Astrophys.Lett., {\bf 14} (1973) 11.

\bi{ll} Landau L.D., Lifshitz E.M. Mechanics.  Pergamon Press, 1953.

\bi{ksf} Kuzovlev Yu.E., Soboleva T.K., Filippov A.E. (Sov.Phys.)  JETP,
{\bf 76}  (1993), 858.

\bi{bos} A.Bosma, Astron.J.,{\bf 86} (1981), 1825.

\bi{rubin} V.C.Rubin, W.K.Ford and N.Thonnard, Astrophys.J., {\bf 238}
(1980), 471.

\bi{arp} Arp H., Atlas of peculiar galaxies, Pasadena, 1966.

\bi{vac} Vacouleurs G. and Vacouleurs A. Reference catalogue of  bright
galaxies. 2-nd edition. Austin, 1976.

\bi{cher1} A.V.Ivanov, E.A.Filistov and A.D.Chernin, Astronomich Zh.,
{\bf 72} (1995) 416 (in Russian).

\eb

\end{document}